\documentclass[conference,letterpaper]{IEEEtran}
\usepackage{cite}
\usepackage{amsmath,amssymb,amsfonts}
\usepackage{algorithmic}
\usepackage{graphicx}
\usepackage{textcomp}
\def\BibTeX{{\rm B\kern-.05em{\sc i\kern-.025em b}\kern-.08em
    T\kern-.1667em\lower.7ex\hbox{E}\kern-.125emX}}

\usepackage{comment}
\usepackage{capt-of}
\usepackage[flushleft]{threeparttable}
\usepackage{array}
\usepackage{url}
\usepackage[bookmarks=false]{hyperref}
\usepackage{bm}
\usepackage{soul}

\usepackage{multirow}
\usepackage{stfloats}

\usepackage[table]{xcolor}
\usepackage{color,soul}
\soulregister\cite7
\soulregister\ref7
\soulregister\pageref7

\newcommand\blfootnote[1]{%
  \begingroup
  \renewcommand\thefootnote{}\footnote{#1}%
  \addtocounter{footnote}{-1}%
  \endgroup
}
\usepackage{enumitem}

\begin{document}

\title{Automated Pupillary Light Reflex Test on a Portable Platform \\
{}
}

\author{\IEEEauthorblockN{Dogancan Temel, Melvin J. Mathew, and Ghassan AlRegib \thanks{This project is sponsored by Georgia Research Alliance (GRA)} }
\IEEEauthorblockA{\textit{School of Electrical and Computer Engineering} \\
\textit{Georgia Institute of Technology}\\
Atlanta, GA, USA \\
\{cantemel, mmathew31, alregib\}@gatech.edu}
\and
\IEEEauthorblockN{Yousuf M. Khalifa}
\IEEEauthorblockA{\textit{School of Medicine, Ophthalmology} \\
\textit{Emory University}\\
Atlanta, GA, USA \\
yousuf.khalifa@emoryhealthcare.org}
}

\twocolumn[{%
\vspace{40mm}
{ \large
\begin{itemize}[leftmargin=2.5cm, align=parleft, labelsep=2.0cm, itemsep=4ex,]

\item[\textbf{Citation}]{D. Temel, M. J. Mathew, G. AlRegib and Y. M. Khalifa, "Automated Pupillary Light Reflex Test on a Portable Platform," International Symposium on Medical Robotics (ISMR), Atlanta, GA, USA, 2019, pp. 1-7.}

\item[\textbf{DOI}]{https://doi.org/10.1109/ISMR.2019.8710182}

\item[\textbf{Review}]{Date added to IEEE Xplore: 09 May  2019}

\item[\textbf{Bib}]  {@INPROCEEDINGS\{Temel2019\_ISMR,\\ 
author=\{D. Temel and M. J. Mathew and G. AlRegib and Y. M. Khalifa\},\\ 
booktitle=\{2019 International Symposium on Medical Robotics (ISMR)\},\\
title=\{Automated Pupillary Light Reflex Test on a Portable Platform\},\\ 
year=\{2019\},\\ 
pages=\{1-7\},\\ 
doi=\{10.1109/ISMR.2019.8710182\},\\ 
month=\{April\},\} }

\item[\textbf{Copyright}]{\textcopyright 2019 IEEE. Personal use of this material is permitted. Permission from IEEE must be obtained for all other uses, in any current or future media, including reprinting/republishing this material for advertising or promotional purposes,
creating new collective works, for resale or redistribution to servers or lists, or reuse of any copyrighted component
of this work in other works. }

\item[\textbf{Contact}]{\href{mailto:alregib@gatech.edu}{alregib@gatech.edu}~~~~~~~\url{https://ghassanalregib.com/}\\ \href{mailto:dcantemel@gmail.com}{dcantemel@gmail.com}~~~~~~~\url{http://cantemel.com/}}
\end{itemize}
\thispagestyle{empty}
\newpage
\clearpage
\setcounter{page}{1}
}
}]

\twocolumn[{%
\renewcommand\twocolumn[1][]{#1}%

\maketitle

\vspace{-8mm}
\centering
\includegraphics[width=0.32\linewidth]{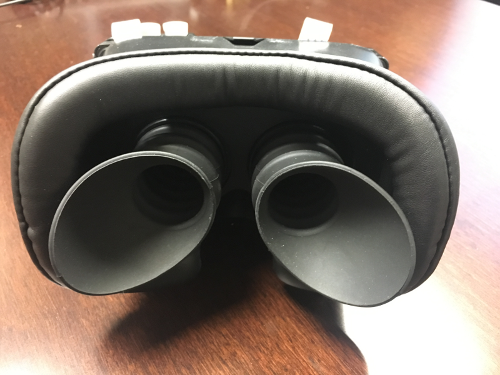}
\includegraphics[width=0.32\linewidth]{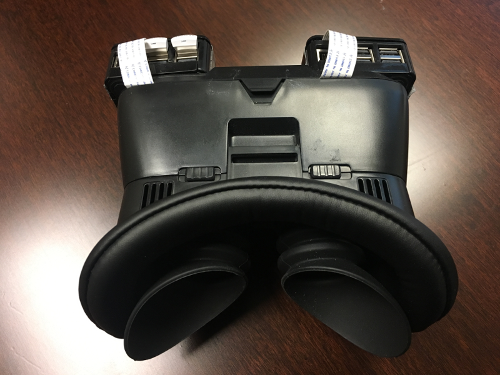}
\includegraphics[width=0.32\linewidth]{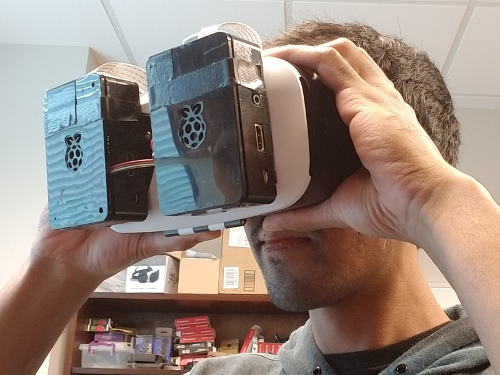}
\captionof{figure}{\texttt{Lab-on-a-headset} platform: Front view, top view, and side view while in use, respectively. }
\label{fig:device}
\vspace{2mm}
}]

\IEEEpubidadjcol

\begin{abstract}
In this paper, we introduce a portable eye imaging device denoted as \texttt{lab-on-a-headset}, which can automatically perform a swinging flashlight test. We utilized this device in a clinical study to obtain high-resolution recordings of eyes while they are exposed to a varying light stimuli. Half of the participants had relative afferent pupillary defect (RAPD) while the other half was a control group. In case of positive RAPD, patient’s pupils constrict less or do not constrict when light stimuli swings from the unaffected eye to the affected eye. To automatically diagnose RAPD, we propose an algorithm based on pupil localization, pupil size measurement, and pupil size comparison of right and left eye during the light reflex test. We validate the algorithmic performance over a dataset obtained from 22 subjects and show that proposed algorithm can achieve a sensitivity of 93.8\% and a specificity of 87.5\%.\blfootnote{*This study is supported by the Georgia Research Alliance  based in Atlanta, Georgia. The purpose of the Georgia Research Alliance support is to explore the commercialization of the technology being studied. The authors of this paper,  Georgia Institute of Technology and Emory University are entitled to royalties related to this research in case of the commercialization of the developed technology.}

\end{abstract}


\section{Introduction}
The term \textit{robot} is introduced by the Czech author Karel \v Capek in his science fiction play \textit{Rossum's Universal Robots} published in 1920 \cite{Capek1920}. The origin of the word is from the Czech word \textit{robota}, which means \textit{tedious and  monotonous routine work}. After more than half a century from the emergence of the word, the field of medicine welcomed its first robot in 1984 with \textit{arthrobot}, which was utilized to position knee joints of patients for a knee replacement surgery \cite{Moustris2011}. After the FDA approval in 2000, medical experts started to use the da Vinci robotic surgical system to perform operations through small incisions with the guidance of its high resolution vision system and display \cite{Hockstein2007}. In 2016, University of Oxford surgeons performed the world's first operation inside the eye using a robot \cite{Edwards2018}. In addition to the surgical automation and assistance systems in ophthalmology, there has been a significant advancement in imaging and diagnosis systems including but not limited to automated detection of diabetic retinopathy in fundus photographs \cite{Gulshan2016} and automated referral recommendation based on OCT images \cite{DeFauw2018}. Advancements in algorithmic analysis not only affected research community but also started to transform clinical practice. On April 11, 2018, the U.S. Food and Drug Administration (FDA) authorized the first device for marketing that can provide a screening decision without the need for a clinician. Authorized device can detect diabetic retinopathy in adults if the condition level is mild or greater \cite{FDA2018}.

\begin{figure*}[h!]
\begin{minipage}[b]{\linewidth}
  \centering
\includegraphics[width=\textwidth]{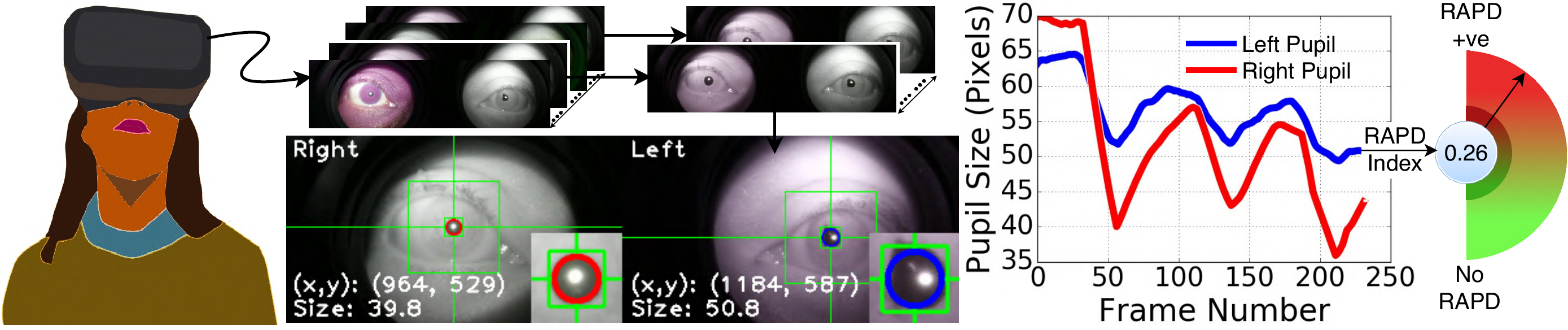}
\end{minipage}
\vspace{0.1cm}
\centering
\vspace{-3 mm}
\caption{RAPD detection framework: We display a light sequence to subjects with \texttt{lab-on-a-headset} and capture their pupillary light reflex synchronously. Then, we measure the dissimilarity between pupillary light reflexes to obtain an index, which can be used to identify patients with RAPD condition.}
\label{fig:rapd_framework}
\vspace{-5 mm}
\end{figure*}

Aforementioned ophthalmology-related studies \cite{Gulshan2016,DeFauw2018,FDA2018}
focused on imaging the back of the eye and the cross-section of retinas to perform frame-level anomaly detection. However, in this study, we need to image the surface of the eye and utilize sequential information to detect abnormalities in pupillary light reflex, which can be an early indicator of ocular and neurological diseases that can eventually lead to vision loss \cite{hall2018}. Specifically, we focus on relative afferent pupillary defect (RAPD), which is based on the difference between the light reflex of the pupils. In healthy subjects, pupils should constrict equally when they are exposed to light and dilate equally when there is no light stimuli. In case of positive RAPD, patient's pupils do not constrict or constrict less when light stimuli swings from the unaffected eye to the affected eye \cite{Broadway2012}. Currently, RAPD assessment is performed by a clinician who asks patients to fixate on a distant point in a relatively dark environment, swings the flashlight between the eyes of the patient, and observe the relative pupil size change. Although subjective RAPD assessment is practical in terms of requiring only a flashlight, variations in test setup and subjective opinion can affect the reliability of the test results. Even the type of the light source can significantly affect the RAPD assessment performance as shown in a study conducted with experienced nurses \cite{Omburo2017}. Moreover, subtle abnormalities are not easy to detect subjectively in case of dark irises, small or poorly reactive pupils \cite{Loewenfeld1971,Kawasaki1995}.

To standardize testing environment and eliminate subjectivity, we introduce a portable eye imaging and light reflex test device denoted as \texttt{lab-on-a-headset} as shown in Fig.~\ref{fig:device}. The introduced device can be considered as a medical imaging device that can perform \textit{tedious and monotonous routine work}, which corresponds to swinging flashlight test in this study as shown in Fig.~\ref{fig:rapd_framework}. The remaining of this paper is organized as follows: We analyze existing pupil dataset studies in Section \ref{sec:related} and describe the introduced dataset in Section \ref{sec:rapd_dataset}. We introduce an RAPD detection algorithm in Section \ref{sec:rapd_algorithm} and report the results in Section \ref{sec:rapd_experiments}. Finally, we conclude our work in Section \ref{sec:conc}.

\section{Related Work}
\label{sec:related}
There are numerous studies in the literature that investigate the pupillary light reflex of subjects based on off-the-shelf digital pupilometers \cite{Volpe2000,Miki2008,Waisbourd2015}. However, these studies cannot be used as a baseline to investigate RAPD detection because they do not disclose utilized algorithms as well as acquired data. Similarly, capturing a new dataset with an off-the-shelf pupilography device is not preferable because of limited control over the stimuli and limited access to raw data. Therefore, we analyzed publicly-available pupil datasets in the literature to understand whether they can be utilized for automated RAPD assessment or not. Kasneci \textit{et al.} \cite{Kasneci2014} conducted a research study to assess the on-road driving performance of subjects during a 40-minute driving task on a specific route. Research study resulted in a dataset of closely captured eye movements, which were utilized to investigate the visual exploration ability of subjects while driving. Half of the participants were healthy control subjects whereas remaining half had Homonymous visual defect or Glaucoma. Sippel \textit{et al.} \cite{Sippel2014} performed a research study to investigate the impact of visual field loss in everyday living activities. Specifically, subjects were asked to collect certain products in a drugstore and their eye movements were recorded during the search task with a closely mounted setup. Half of the participants were healthy control subjects and the reaming half had Binocular Glaucoma.

Fuhl \textit{et al.} \cite{Fuhl2015_ExCuSe} investigated the robustness of pupil detection in real-world challenging scenarios including varying illumination conditions, reflection on eyeglasses and contact lenses. In addition to the dataset introduced by Swirski \textit{et al.} \cite{Swirski2012}, Fuhl \textit{et al.} \cite{Fuhl2015_ExCuSe} evaluated their algorithm over nine image sets from on-road experiments \cite{Kasneci2014} and eight image sets from supermarket experiments \cite{Sippel2014}. Fuhl \textit{et al.} \cite{Fuhl2016_ElSe} extended the dataset in \cite{Fuhl2015_ExCuSe} with more challenging scenarios to test pupil detection performance in real-world environments. Five new image sets were obtained from on-road experiments \cite{Kasneci2014} with motion blur, reflections, and low pupil contrast. And two new image sets were obtained indoor from Asian subjects whose eyelids and eyelashes partially covered pupils along with reflections in certain images. Fuhl \textit{et al.} (Microscope) \cite{Fuhl2016_Microscope} focused on pupil detection tailored for microscope images and introduced a pupil dataset whose images were obtained with an unmodified microscope ocular. Channeling conditions in the microscope dataset include blur, exposure, contrast, irregular pupil shape, light gradient, and reflections. Tonsen \textit{et al.} \cite{Tonsen2016_LPW} introduced the labeled pupils in the wild (LPW) dataset to study pupil detection in unconstrained environments, which includes indoor/outdoor environments, participants wearing glasses and eye make-up from different ethnicities with variable skin tones, eye colours, and face shapes. 


Existing pupil datasets can not be directly used for pupillary light reflex assessment for three main reasons. First, majority of existing datasets are based on images, which lack the sequential information necessary to assess light reflex. Kasneci \cite{Kasneci2014} and Sippel \cite{Sippel2014} were acquired as video sequences but only static images were provided in related studies \cite{Fuhl2015_ExCuSe,Fuhl2016_ElSe}. Second, limited control over acquisition conditions makes it impossible to assess the reflex based on solely light stimuli because of simultaneous exposure to varying conditions. Third, there is no medical metadata corresponding to the RAPD condition. Because of these limitations, we decided to obtain a new dataset for pupillary light reflex assessment. An intuitive option would be capturing pupil data with off-the-shelf portable eye trackers \cite{Kassner2014,Tobii2010} that are commonly used for related research studies. Even though eye trackers are optimized for gaze and eye tracking, their design leads to inherent limitations for measuring constriction and dilation motion. Imaging sensors of the eye trackers are located around the eyes and have an indirect view of the pupils, which makes it more challenging to accurately measure pupil size. Moreover, integrating an automated light stimulus with an off-the-shelf eye tracker is not straightforward because of the black box nature and limited control over the eye tracker. Thus, we decided to develop a custom acquisition and test platform that can be used in a clinical study to obtain a RAPD dataset.

\section{RAPD Dataset}
\label{sec:rapd_dataset}
\begin{figure}[h]
\centering
\begin{minipage}[b]{0.32\linewidth}
  \centering
\includegraphics[width=\textwidth]{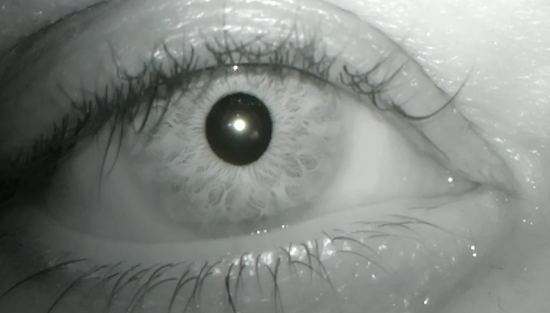}
\end{minipage}
\begin{minipage}[b]{0.32\linewidth}
  \centering
\includegraphics[width=\linewidth]{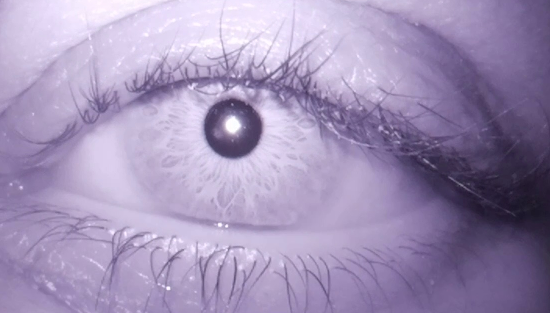}
\end{minipage}
\begin{minipage}[b]{0.32\linewidth}
  \centering
\includegraphics[width=\textwidth]{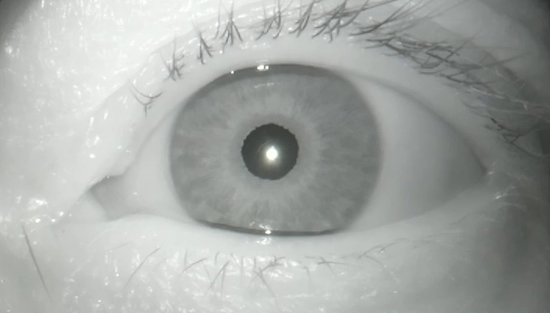}
\end{minipage}

\begin{minipage}[b]{0.32\linewidth}
  \centering
\includegraphics[width=\textwidth]{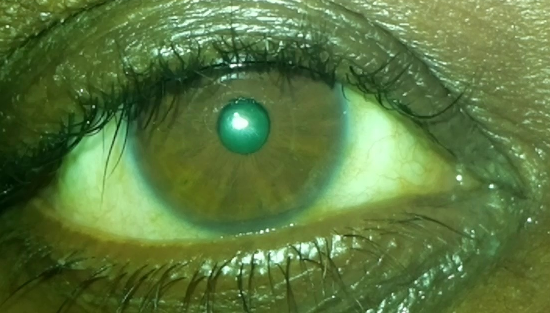}
\end{minipage}
\begin{minipage}[b]{0.32\linewidth}
  \centering
\includegraphics[width=\linewidth]{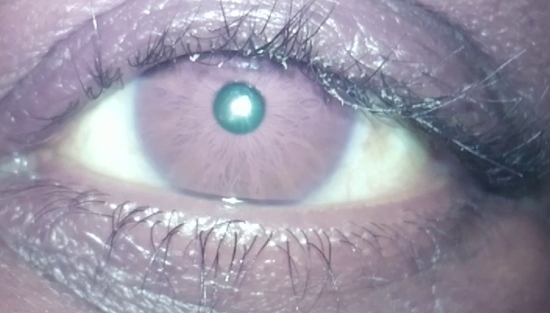}
\end{minipage}
\begin{minipage}[b]{0.32\linewidth}
  \centering
\includegraphics[width=\textwidth]{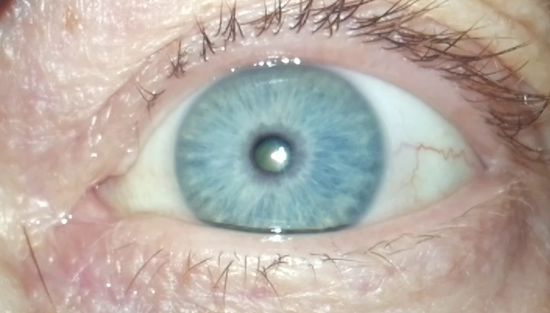}
\end{minipage}

\centering
\caption{Sample pupil images obtained with \texttt{lab-on-a-headset}.}
\label{fig:sample_pupils}
\end{figure}

We utilized \texttt{lab-on-a-headset} \cite{Temel2018_RAPD_Patent,Temel2019_RAPD_Patent} to perform automated light rexlex test and generated our RAPD dataset. \texttt{Lab-on-a-headset} is an ultra portable device with on-board processing capability, which enabled us to capture and preview patient videos during the clinical study. We predefined the test sequence by connecting to the headset and utilizing the graphical user interface that provides full control over the test stimuli. Tested subjects were stimulated with the automated light sequences and their pupillary reactions were recorded simultaneously as high definition streams. In Fig.~\ref{fig:sample_pupils}, we show sample images captured with the introduced headset. We used the infrared frames to assess relative afferent pupillary defect in this study.

\begin{figure*}[b!]
\begin{minipage}[b]{0.98\linewidth}
  \centering
\includegraphics[width=\textwidth]{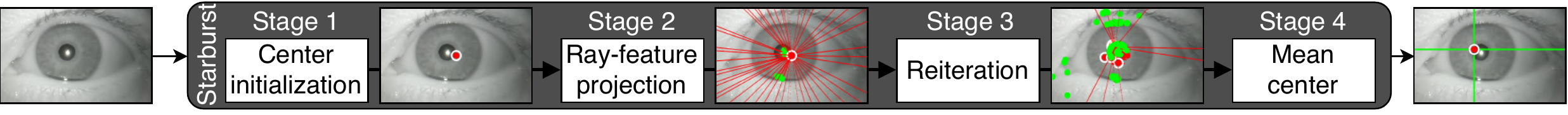}
\end{minipage}
\centering
\caption{Pupil detection pipeline based on Starburst algorithm.}
\label{fig:starburst}
\end{figure*}

We obtained approvals from the Institutional Review Board committees of Emory University and Georgia Institute of Technology. RAPD conditions of the subject were determined by clinicians with manual swinging flashlight test as well as neutral density filter test. The demographics of clinical subjects are summarized in Table~\ref{tab:dataset_rapd}. The RAPD dataset used in study included 22 subjects, half of which correspond to a control group without RAPD whereas the other half have positive RAPD. Four out of ten males and seven out of twelve females have positive RAPD. Average age of the participants is around 52 years for no RAPD subjects and 56 years for RAPD positive subjects with a standard deviation of approximately 14 and 10, respectively.   

\begin{table}[htbp!]
\centering
\caption{Subject statistics in the clinical study. }
\label{tab:dataset_rapd}
\begin{tabular}{ccc}
\hline
\textbf{} & \textbf{No RAPD} & \textbf{RAPD +ve} \\ \hline

\textbf{\begin{tabular}[c]{@{}c@{}}\# Subjects (\% of total) \end{tabular}}
& \begin{tabular}[c]{@{}c@{}} 11 (50.00\%) \end{tabular}
& \begin{tabular}[c]{@{}c@{}} 11 (50.00\%) \end{tabular}
 \\ \hline

\textbf{\begin{tabular}[c]{@{}c@{}}\# Males (\% of group) \end{tabular}}
& \begin{tabular}[c]{@{}c@{}} 6 (54.55\%) \end{tabular}
& \begin{tabular}[c]{@{}c@{}} 4 (36.36\%) \end{tabular}
 \\ \hline

\textbf{\begin{tabular}[c]{@{}c@{}}\# Females (\% of group) \end{tabular}}
& \begin{tabular}[c]{@{}c@{}} 5 (45.45\%) \end{tabular}
& \begin{tabular}[c]{@{}c@{}} 7 (63.64\%) \end{tabular}
 \\ \hline

\textbf{\begin{tabular}[c]{@{}c@{}}Age (mean $\pm$ $\sigma$, in years) \end{tabular}}
& \begin{tabular}[c]{@{}c@{}} 52.27 $\pm$ 14.48 \end{tabular}
& \begin{tabular}[c]{@{}c@{}} 56.73 $\pm$ 10.53 \end{tabular}
 \\ \hline
\end{tabular}
\end{table}

\section{RAPD Detection Algorithm}
\label{sec:rapd_algorithm}

\subsection{Pupil Detection}
\label{subsec:pupil_localization}
\begin{figure}[h!]
\begin{minipage}[b]{0.98\linewidth}
  \centering
\includegraphics[width=\textwidth]{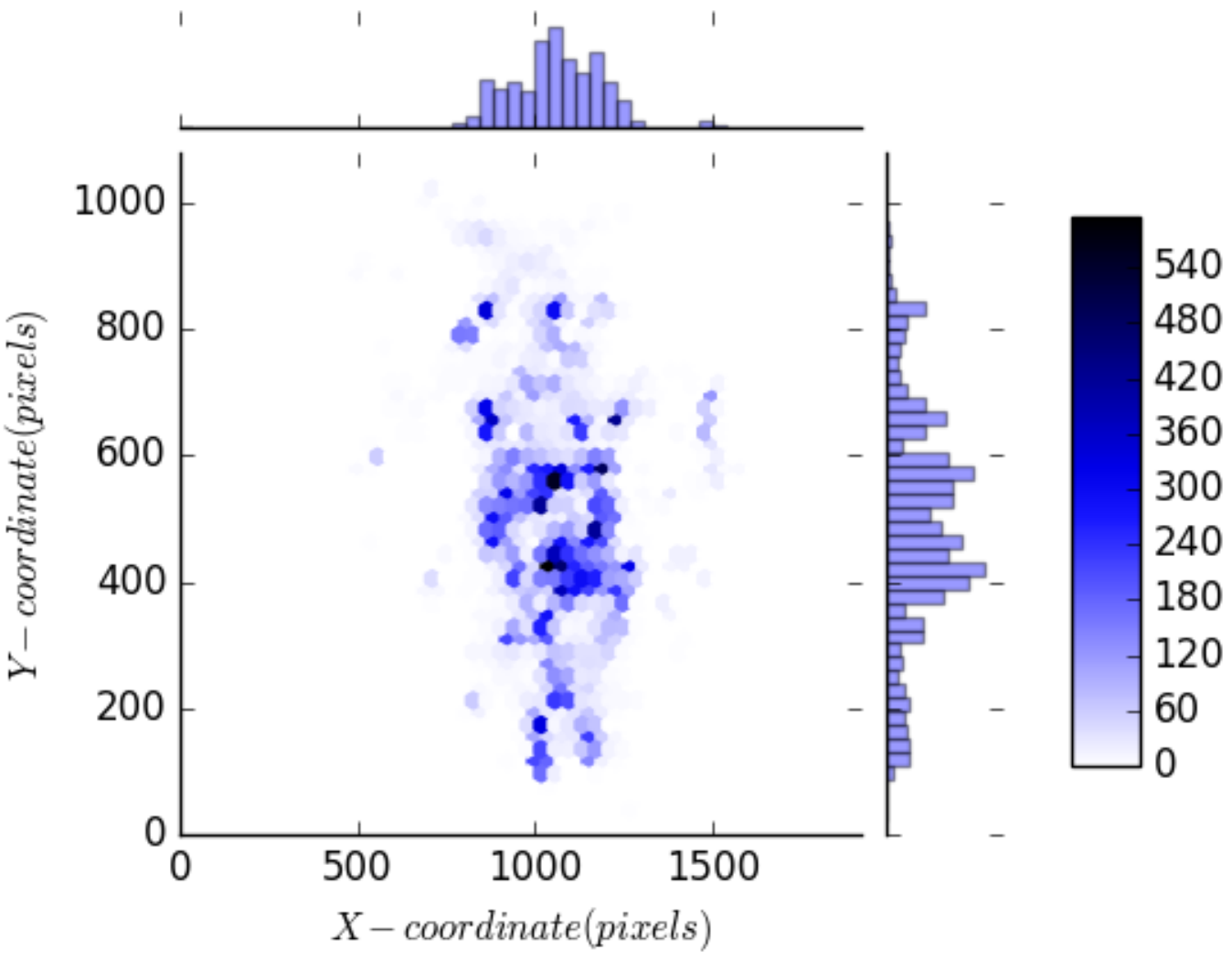}
\end{minipage}
\vspace{0.1cm}
\centering
\vspace{-5 mm}
\caption{Heat map of the estimated pupil centers in the test sequences. X-axis corresponds to horizontal pupil location, y-axis corresponds to vertical pupil location, and the color represents the density, which switches from white to dark blue as more pupils are detected over the same region. Distrubtion of pupil centers are also shown on the right side and the top of the heat map.}
\label{fig:pupil_distribution}
\end{figure}

We utilize the Starburst algorithm  for pupil localization \cite{Li2005_Starburst}. Specifically, we implemented our own version based on the details in the paper \cite{Li2005_Starburst} and the website \cite{starburst_code}. We can group the main components in the algorithm into four stages as (1) initialization of the pupil center, (2) ray-feature projection, (3) reiteration, and (4) mean calculation as shown in Fig.\ref{fig:starburst}. Based on the \texttt{lab-on-a-headset} setup, pupils are usually located within the central region of the video frames as shown in Fig.~\ref{fig:pupil_distribution}. Therefore, we initialize the pupil center as the center of the video frames.

\begin{figure}[htbp!]
\centering
\begin{minipage}[b]{0.32\linewidth}
  \centering
\includegraphics[width=\textwidth]{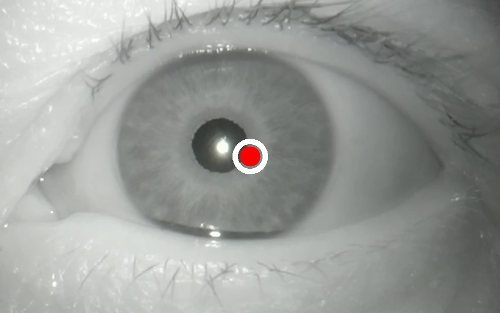}
  \centerline{\footnotesize{(a)}}
\end{minipage}
\begin{minipage}[b]{0.32\linewidth}
  \centering
\includegraphics[width=\linewidth]{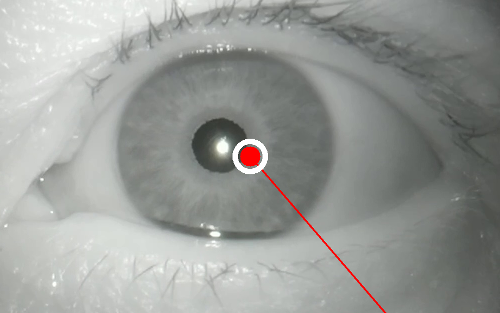}
  \centerline{\footnotesize{(b)} }
\end{minipage}
\begin{minipage}[b]{0.32\linewidth}
  \centering
\includegraphics[width=\textwidth]{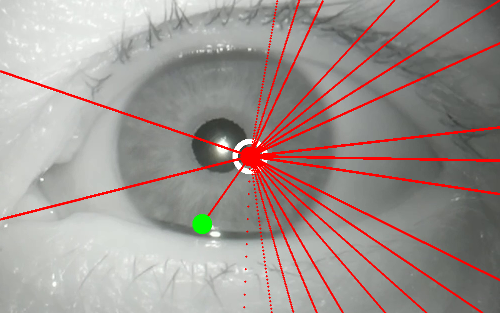}
  \centerline{\footnotesize{(c)}}
\end{minipage}

\begin{minipage}[b]{0.32\linewidth}
  \centering
\includegraphics[width=\textwidth]{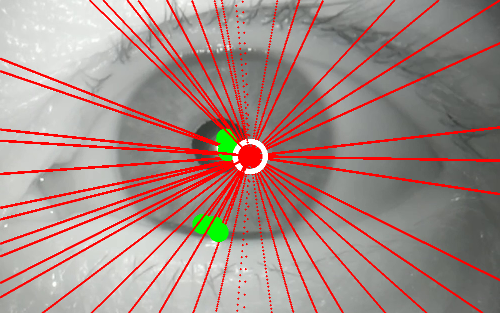}
  \centerline{\footnotesize{(d)}}
\end{minipage}
\begin{minipage}[b]{0.32\linewidth}
  \centering
\includegraphics[width=\linewidth]{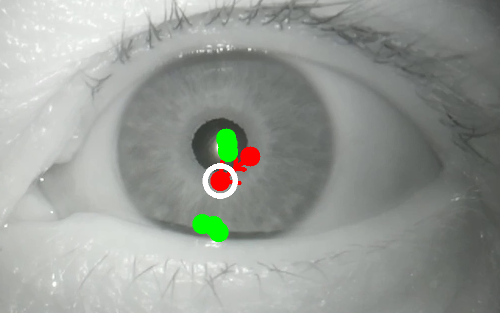}
  \centerline{\footnotesize{(e)} }
\end{minipage}
\begin{minipage}[b]{0.32\linewidth}
  \centering
\includegraphics[width=\textwidth]{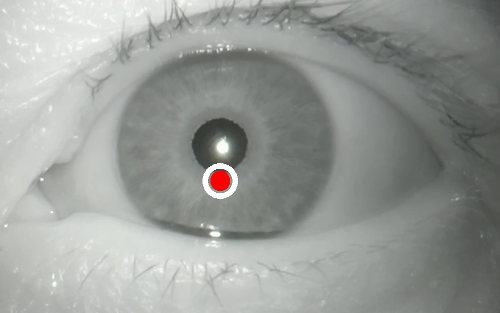}
  \centerline{\footnotesize{(f)}}
\end{minipage}

\begin{minipage}[b]{0.32\linewidth}
  \centering
\includegraphics[width=\textwidth]{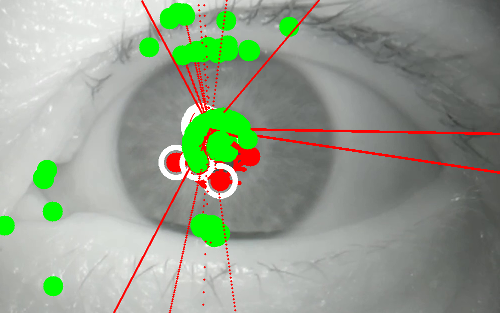}
  \centerline{\footnotesize{(g)}}
\end{minipage}
\begin{minipage}[b]{0.32\linewidth}
  \centering
\includegraphics[width=\linewidth]{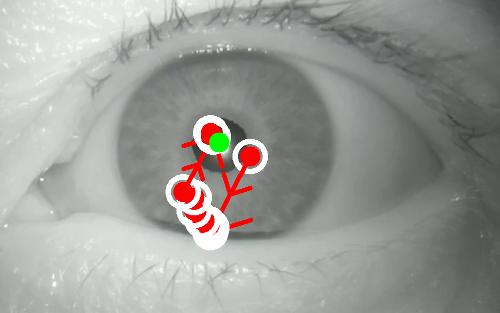}
  \centerline{\footnotesize{(h)} }
\end{minipage}
\begin{minipage}[b]{0.32\linewidth}
  \centering
\includegraphics[width=\textwidth]{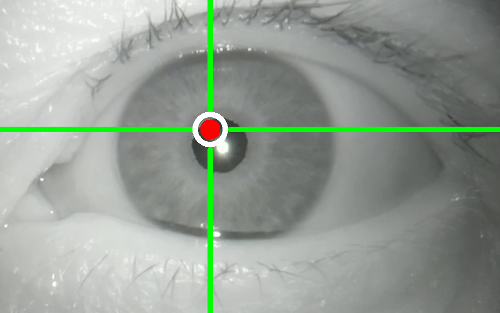}
  \centerline{\footnotesize{(i)}}
\end{minipage}

\centering
\caption{Visualization of the stages in the Starburst algorithm including center initialization (a), ray-feature projection (b-f), reiteration (g-h), and averaging of the feature centers (i).}
\label{fig:starburst_detailed}
\end{figure}

In Stage 2, rays are projected outwardly from the initialized pupil center to determine feature points, which are used to update the pupil center. We provide a detailed illustration of the progression of Stage 2 in Fig.~\ref{fig:starburst_detailed}. We can observe the initialized pupil center in Fig.~\ref{fig:starburst_detailed}(a) and first projected ray from the pupil center in Fig.~\ref{fig:starburst_detailed}(b). We then compare the gradient value of every point along this ray against a gradient threshold value. Ray projection continues until either a valid feature point is determined or until the ray reaches the border of the image frame. Gradient threshold test does not result in any detected feature points. Thus, the same process is repeated for rays projected in all directions as shown in Fig.~\ref{fig:starburst_detailed}(c) in which the green point represents a valid feature point. After determining all feature points on all rays, the mean of the feature points is used to update the pupil center. We show this in Fig.~\ref{fig:starburst_detailed}(d)-(f). In Stage 3, we repeat the steps in Stage 2 using the updated pupil center and a new threshold value to determine the next updated pupil center. This is repeated until all gradient values are swept through, ranging from 255 to 0, which leads to feature points shown in Fig.~\ref{fig:starburst_detailed}(g) and estimated pupil centers in Fig.~\ref{fig:starburst_detailed}(h). In Stage 4, we calculate the mean of feature points to determine the final pupil location as shown in Fig.~\ref{fig:starburst_detailed}(i).

\subsection{Pupil Size Measurement}
\label{subsec:pupil_measurement}
\begin{figure}[htbp!]
\begin{minipage}[b]{\linewidth}
  \centering
\includegraphics[width=\textwidth]{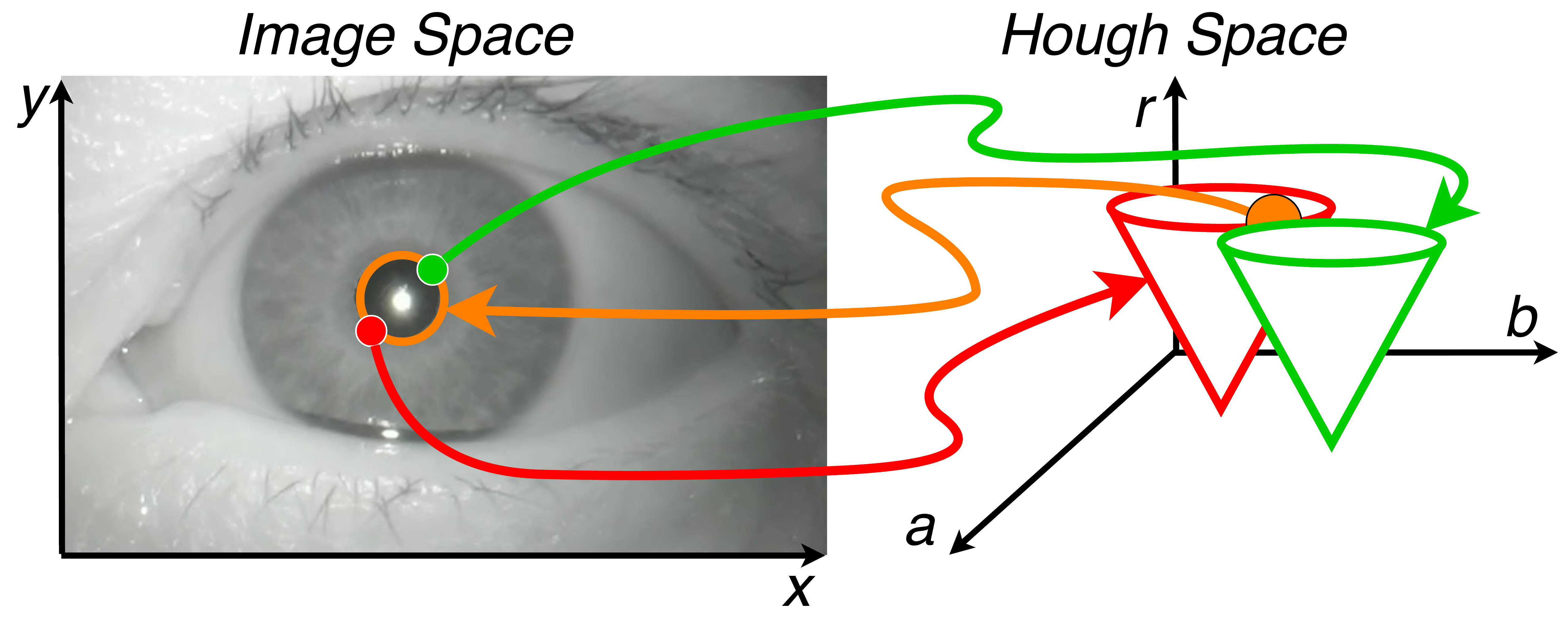}
\end{minipage}
\centering
\vspace{-3 mm}
\caption{Circular Hough Transform from the image plane to the Hough space.}
\label{fig:cht_hough_space}
\vspace{-3 mm}
\end{figure}

We can consider pupil as a circular region and utilize the Circular Hough Transform (CHT) to measure the pupil size. We use CHT to transform the pixels in the image plane (x,y) into right circular cones in the Hough space. In Fig.\ref{fig:cht_hough_space}, we provide an example in which red and orange points over the pupil circumference are transformed into Hough domain. We can observe that transformed points correspond to cones in the Hough domain and the intersection of the cones lead to the parameters of the pupil in the spatial domain. 

\begin{figure}[htbp!]
  \centering
\vspace{-3mm}
\begin{minipage}[b]{0.48\linewidth}
  \centering
\includegraphics[width=\textwidth]{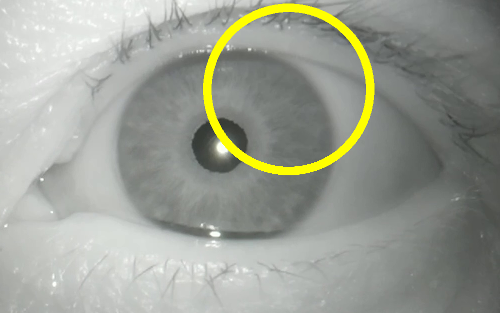}
  \vspace{0.01cm}
  \centerline{\footnotesize{(a) Original image}}
  \vspace{0.03 cm}
\end{minipage}
\vspace{0.1cm}
\begin{minipage}[b]{0.48\linewidth}
  \centering
\includegraphics[width=\textwidth]{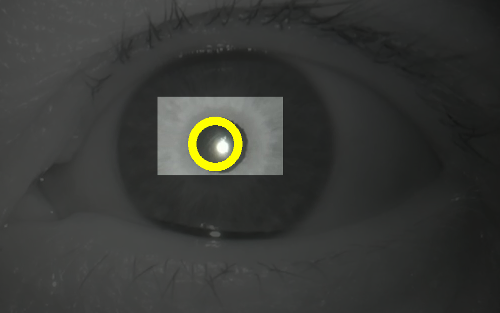}
  \vspace{0.01cm}
  \centerline{\footnotesize{(b) Highlighted cropped region }}
  \vspace{0.03 cm}
\end{minipage}

\centering
\vspace{-4 mm}
\caption{Pupil size measurement over entire image versus cropped image.}
\label{fig:crop_example}
\end{figure}

We need to determine a minimum value and a maximum value for the search range of the circles. In this study, we set the max search range as the entire image to consider pupil size variation among subjects and min as the 5\% of the max value to eliminate smaller regions that can correspond to glint or other pointy regions. We perform a parameter sweep to determine the accumulator threshold, the accumulator array size, and the canny threshold. The accumulator array is used to determine the point of intersection between all cones in the Hough parameter space. The accumulator threshold is used to determine the number of votes required for a point of intersection to be considered as a circle in the image plane. The canny threshold is used to transform the image into a binary map and separate the brighter regions from the dark region, which should correspond to pupils. We perform pupil size measurement over entire pupil images (Fig.~\ref{fig:crop_example}(a)) as well as over images cropped around estimated pupil center (Fig.~\ref{fig:crop_example}(b)). Specifically, we crop the image around the estimated pupil location to quarter of input image size and measure the pupil size over the cropped region. We can observe that surrounding structures in the eye image corrupt the pupil measurement whereas we obtain a more accurate pupil measurement in the cropped scenario.

\subsection{RAPD Assessment}
\begin{figure}[htbp!]
\centering
\begin{minipage}[b]{0.48\linewidth}
  \centering
\includegraphics[width=\textwidth]{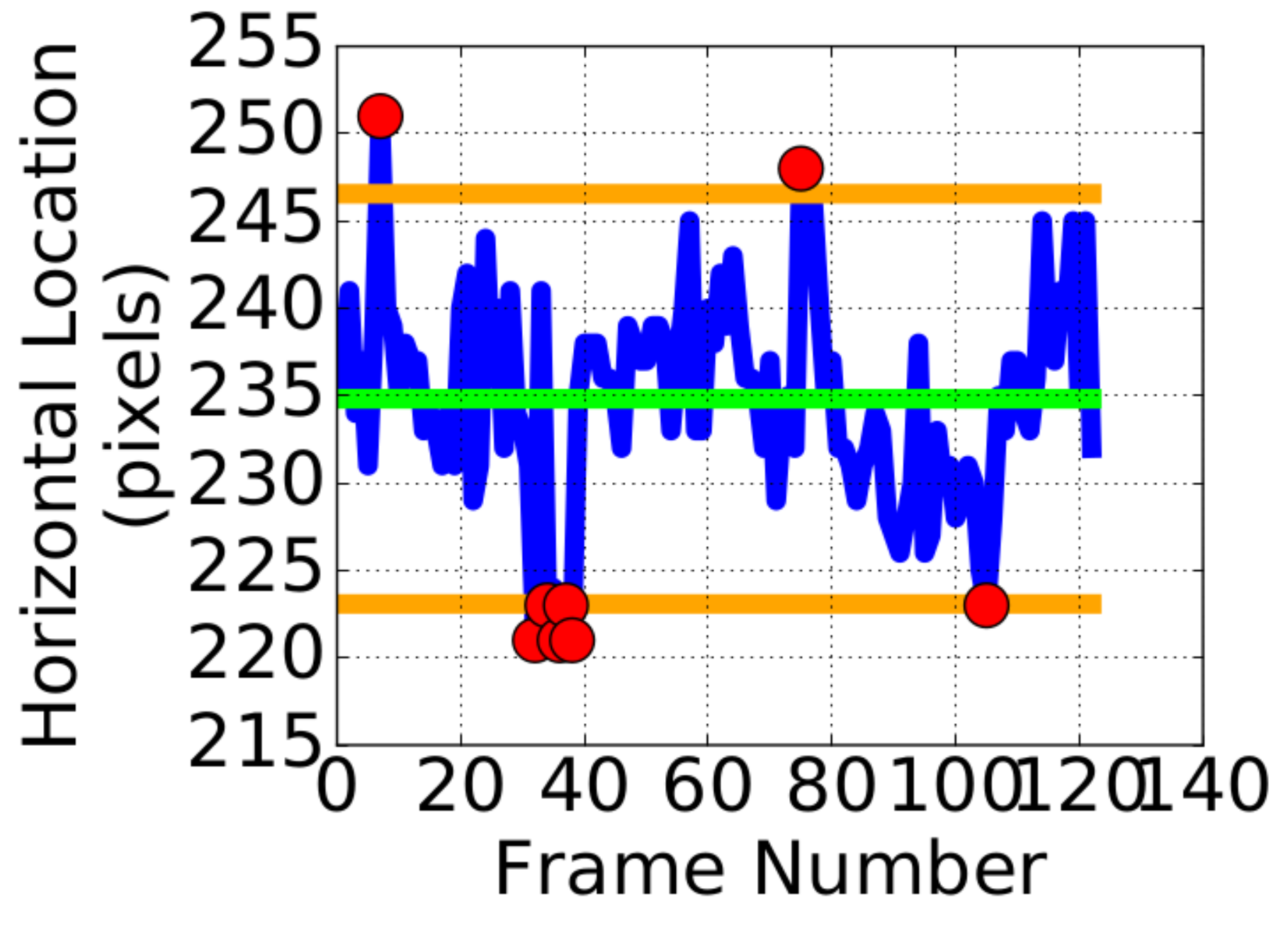}
  \vspace{0.01cm}
  \centerline{\footnotesize{(a) Pupil horizontal motion}}
  \vspace{0.03 cm}
\end{minipage}
\vspace{0.1cm}
\begin{minipage}[b]{0.48\linewidth}
  \centering
\includegraphics[width=\textwidth]{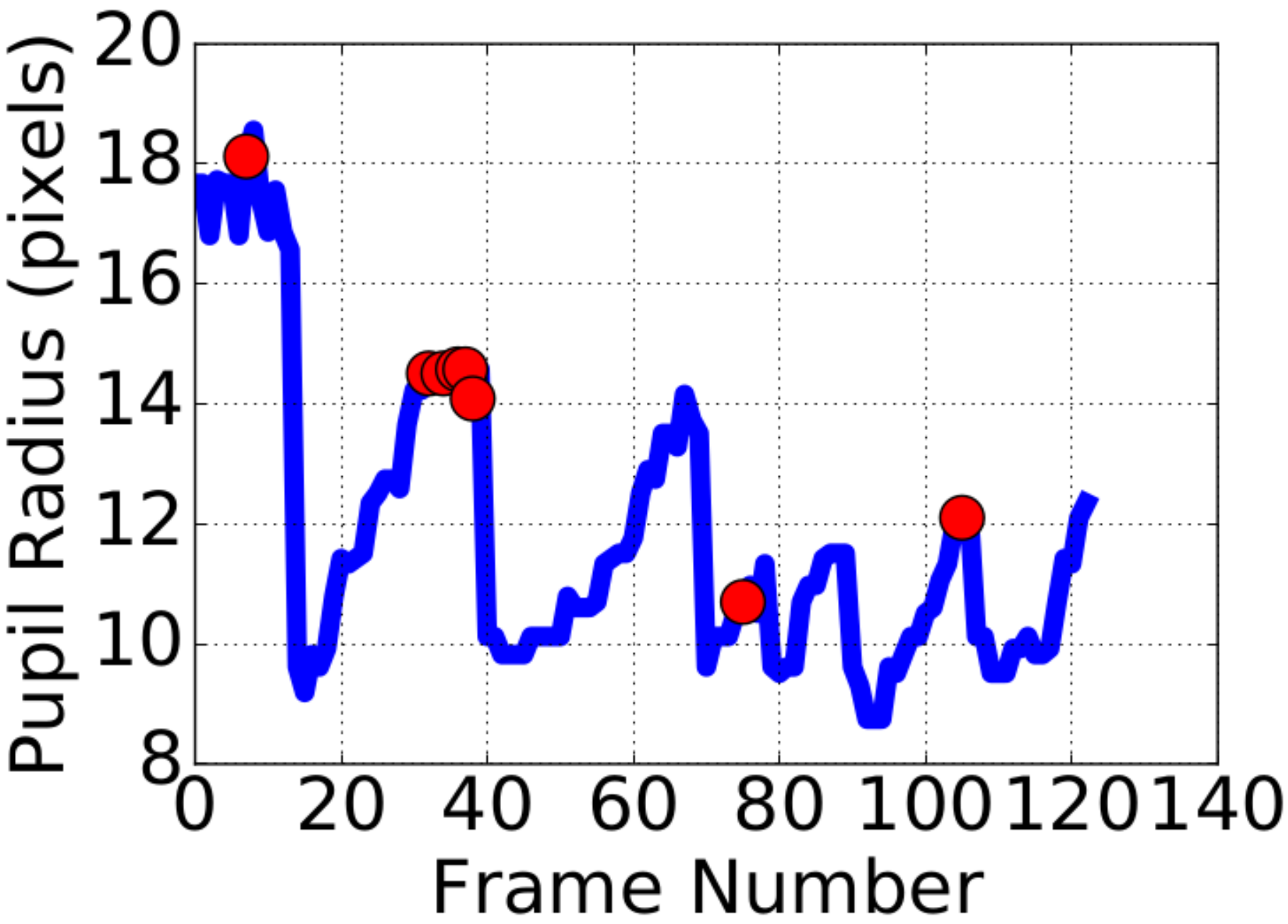}
  \vspace{0.01cm}
  \centerline{\footnotesize{(b) Pupil size measurement}}
  \vspace{0.03 cm}
\end{minipage}

\begin{minipage}[b]{0.48\linewidth}
  \centering
\includegraphics[width=\textwidth]{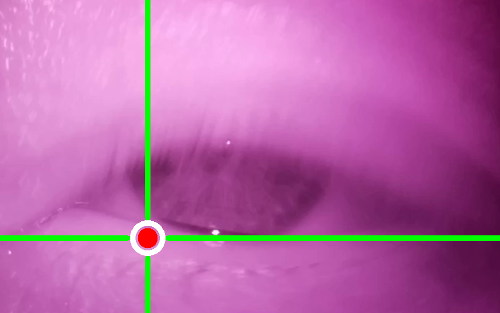}
  \vspace{0.01cm}
  \centerline{\footnotesize{(c) Deviation from mean}}
  \vspace{0.03 cm}
\end{minipage}
\vspace{0.1cm}
\begin{minipage}[b]{0.48\linewidth}
  \centering
\includegraphics[width=\textwidth]{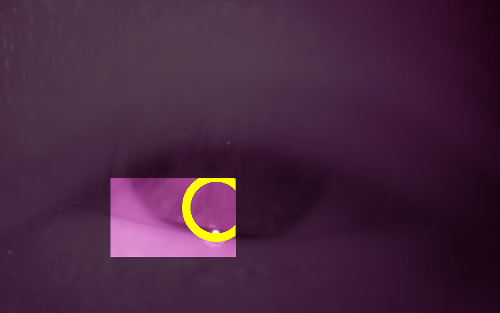}
  \vspace{0.01cm}
  \centerline{\footnotesize{(d) Invalid measurement}}
  \vspace{0.03 cm}
\end{minipage}
\centering
\vspace{-4 mm}
\caption{An erroneous pupil detection example in which false pupil detection can be determined from the horizontal motion of the pupils.}
\vspace{-2 mm}
\label{fig:postprocess_motion}
\end{figure}

Inaccurate variants of pupil size measurement can mislead the RAPD assessment process. In Fig.~\ref{fig:postprocess_motion}, we provide a sample from the conducted swinging flashlight test. Horizontal motion of the pupil is shown in Fig.~\ref{fig:postprocess_motion}(a) in which x-axis corresponds to frame number of the video, y-axis corresponds to the horizontal location of the pupil, and green line indicates the mean horizontal location of the pupil. Based on the preliminary analysis of the test sequences, we observed that horizontal/vertical pupil motion is usually restricted to 5\% of the mean location, which is shown with the orange lines in Fig.~\ref{fig:postprocess_motion}(a). We highlight the data points exceeding the normal motion range with red in Fig.~\ref{fig:postprocess_motion}(a) and provide corresponding pupil size measurements in Fig.~\ref{fig:postprocess_motion}(b).  We show a pupil localization result corresponding to one of these erroneous points in Fig.~\ref{fig:postprocess_motion}(c) along with detected pupil in Fig.~\ref{fig:postprocess_motion}(d). We can observe that even though the radius of the measurement is in a normal range, detected pupil corresponding to erroneous motion is inaccurate. Therefore, we filter out the detection results that exceed pupil motion limit both horizontally and vertically before comparing right and left pupillary light reflex. 

\begin{figure*}[b!]
\begin{minipage}[]{0.34\linewidth}
  \centering
\includegraphics[width=\textwidth]{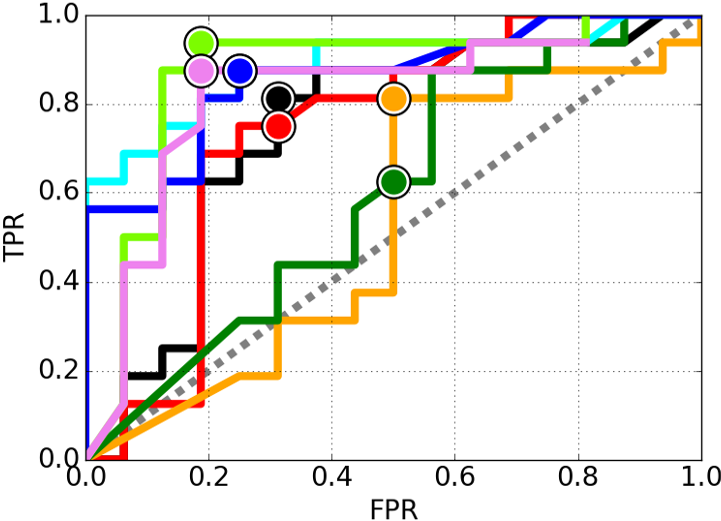}
  \vspace{0.01cm}
  \centerline{\footnotesize{(a)Dissimilarity: 1-SRCC}}
  \vspace{0.03 cm}
\end{minipage}
\vspace{0.1cm}
\begin{minipage}[]{0.34\linewidth}
  \centering
\includegraphics[width=\textwidth]{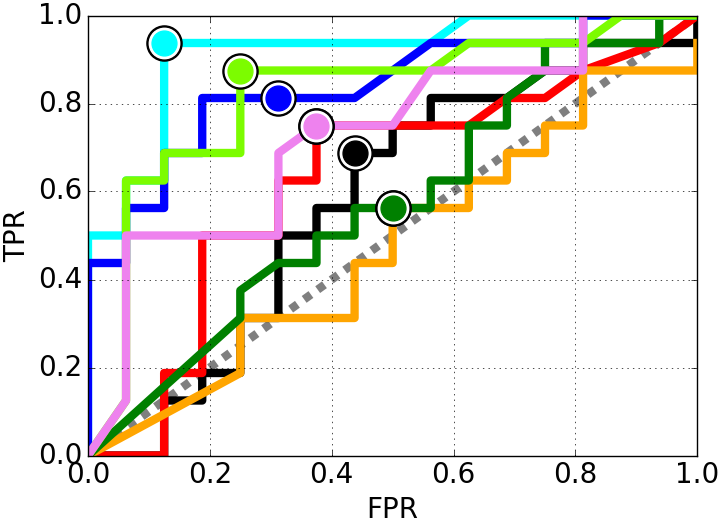}
  \vspace{0.01cm}
  \centerline{\footnotesize{(b)Dissimilarity: 1-PLCC}}
  \vspace{0.03 cm}
\end{minipage}
\vspace{0.1cm}
\begin{minipage}[]{0.28\linewidth}
  \centering
\includegraphics[width=\textwidth]{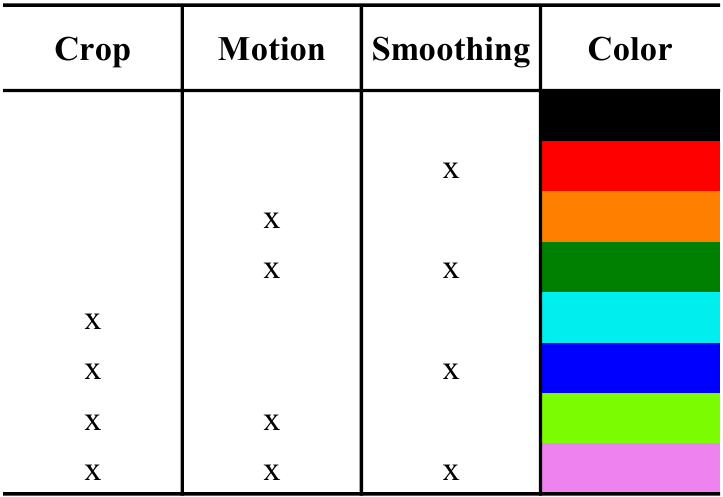}
  \vspace{0.01cm}
  \vspace{0.03 cm}
\end{minipage}
\centering
\caption{Receiver operator curves (ROC) for RAPD detection algorithms. There are 16 algorithm configurations, half of which are based on residual of Spearman correlation (a) as dissimlarity measure and the remaining ones are based on residual of Pearson correlation (b). In each subfigure, there
are 8 ROC curves that correspond to distinct configurations based on cropping, motion-based postprocessing,
and smoothing as shown in the color legend.}
\label{fig:results_ROC}
\end{figure*}

We assess the RAPD condition by measuring the dissimilarity between right and left pupillary reflex. At first we measure the similarity between right and left reflex in terms of  Spearman rank correlation coefficient (SRCC) and Pearson linear correlation coefficient (PLCC). Then, we calculate one minus absolute correlation to obtain dissimilarity. For each dissimilarity measure, we perform a combination of crop, motion-based postprocessing, and smoothing. Crop refers to measuring pupils over cropped images as described in Section~\ref{subsec:pupil_measurement}. Smoothing refers to median filtering the pupil size curves with a window size of three before dissimilarity comparison. Motion-based postprocessing refers to removal of pupil size measurements that correspond to erroneous pupil motion as described in the previous paragraph.

Dissimilarity measures approach zero as the correlation between relative pupil size measurements increases.
Spearman correlation focuses on the monotonic relationship between compared signals whereas Pearson correlation focuses on the linear relationship. When a subject does not have RAPD, pupillary reactions should be almost identical, which would lead to a strong monotonic and linear relationship between compared signals. Therefore, dissimilarity measure should be close to zero as correlation measures are close to one. In case of RAPD, monotonic and linear relationship should be weaker, which would lead to low correlation values and high dissimilarity values.

\section{Experimental Results and Discussion}
\label{sec:rapd_experiments}
\begin{table}[htbp!]
\centering
\caption{Description of the terms used for detection performance.}
\begin{tabular}{c|c}
\hline
\textbf{Term} & \textbf{Description} \\ \hline
Positive ($P$) & Number of RAPD positive subjects \\ 
Negative ($N$) & Number of subjects without RAPD \\ 
True positive ($TP$) & Number of correct RAPD detections \\ 
True negative ($TN$) & Number of  correct no RAPD detections \\ 
False positive ($FP$) & Number of false RAPD detection \\ 
False negative($FN$) & Number of undetected RAPD patients\\ \hline
\end{tabular}
\label{tab:metrics}
\end{table}

\begin{table*}[h]
\centering
\caption{Overall evaluation of RAPD detection algorithms.}
\label{tab:results}
\begin{tabular}{c|c|c|c|c|c|c|c|c|c|c|c}
\hline
\multirow{2}{*}{\textbf{Crop}} & \multirow{2}{*}{\textbf{Motion}} & \multirow{2}{*}{\textbf{Smoothing}} & \multicolumn{2}{c|}{\textbf{Dissimilarity}} & \multirow{2}{*}{\textbf{Sensitivity}} & \multirow{2}{*}{\textbf{Specificity}} & \multirow{2}{*}{\textbf{Precision}} & \multirow{2}{*}{\textbf{AUC}} & \multirow{2}{*}{\bm{$F_{0.5}$}} & \multirow{2}{*}{\bm{$F_1$}} & \multirow{2}{*}{\bm{$F_2$}} \\ \cline{4-5}
 &  &  & \textbf{1-SRCC} & \textbf{1-PLCC} &  &  &  &  &  &  &  \\ \hline \hline
              &                 &                    & x              &                & 81.3\%                          & 68.8\%                   & 72.2\%                 & 0.738        & 1.912           & 0.765         & 0.478         \\
              &                 &                    &                & x              & 68.8\%                          & 56.3\%                   & 61.1\%                 & 0.580        & 1.618           & 0.647         & 0.404         \\
              &                 & x                  & x              &                & 75.0\%                          & 68.8\%                   & 70.6\%                 & 0.738        & 1.818           & 0.727         & 0.455         \\
              &                 & x                  &                & x              & 75.0\%                          & 62.5\%                   & 66.7\%                 & 0.627        & 1.765           & 0.706         & 0.441         \\
              & x               &                    & x              &                & 81.3\%                          & 50.0\%                   & 61.9\%                 & 0.527        & 1.757           & 0.703         & 0.439         \\
              & x               &                    &                & x              & 56.3\%                          & 50.0\%                   & 52.9\%                 & 0.473        & 1.364           & 0.545         & 0.341         \\
              & x               & x                  & x              &                & 62.5\%                          & 50.0\%                   & 55.6\%                 & 0.596        & 1.471           & 0.588         & 0.368         \\
              & x               & x                  &                & x              & 56.3\%                          & 50.0\%                   & 52.9\%                 & 0.570        & 1.364           & 0.545         & 0.341         \\
x             &                 &                    & x              &                & 87.5\%                          & 75.0\%                   & 77.8\%                 & 0.885        & 2.059           & 0.824         & 0.515         \\
\cellcolor{yellow!30}x             & \cellcolor{yellow!30}                & \cellcolor{yellow!30}                   &\cellcolor{yellow!30}                & \cellcolor{yellow!30}x              &\cellcolor{yellow!30} \bf 93.8\%                          &\cellcolor{yellow!30} \bf 87.5\%                   & \cellcolor{yellow!30} \bf 88.2\%                 & \cellcolor{yellow!30} \bf 0.916        & \cellcolor{yellow!30} \bf 2.273           & \cellcolor{yellow!30} \bf 0.909         & \cellcolor{yellow!30} \bf 0.568         \\
x             &                 & x                  & x              &                & 87.5\%                          & 75.0\%                   & 77.8\%                 & 0.861        & 2.059           & 0.824         & 0.515         \\
x             &                 & x                  &                & x              & 81.3\%                          & 68.8\%                   & 72.2\%                 & 0.840        & 1.912           & 0.765         & 0.478         \\
x             & x               &                    & x              &                &\bf 93.8\%                          & 81.3\%                   & 83.3\%                 & 0.676        & 2.206           & 0.882         & 0.551         \\
x             & x               &                    &                & x              & 87.5\%                          & 75.0\%                   & 77.8\%                 & 0.820        & 2.059           & 0.824         & 0.515         \\
x             & x               & x                  & x              &                & 87.5\%                          & 81.3\%                   & 82.4\%                 & 0.635        & 2.121           & 0.848         & 0.530         \\
x             & x               & x                  &                & x              & 75.0\%                          & 62.5\%                   & 66.7\%                 & 0.537        & 1.765           & 0.706         & 0.441         \\ \hline
\end{tabular}
\end{table*}

At first, we analyze the receiver operating characteristics (ROC) curves of RAPD detection algorithms to evaluate their diagnostic capability. We obtain the ROC curves in Fig.~\ref{fig:results_ROC} by plotting the true positive rate ($TPR$) in y-axis versus false positive rate ($FPR$) in x-axis of RAPD detection algorithms while sweeping the classification threshold. We calculate $TPR$ as $TP/P$ and false positive rate as  $FP/N$, which are described in Table~\ref{tab:metrics}. For each algorithm configuration, we determine the optimal operation points that maximize sensitivity and specificity, which are shown with a filled circle over each curve. We calculate $sensitivity$ as $TP/P$ and $specificity$ as $TN/N$. In addition to $sensitivity$ ($recall$) and $specificity$, we evaluate the RAPD detection performance in terms of $precision$ ($TP/(TP+FP)$), $AUC$, and $F~scores$. We obtain $AUC$ values by calculating the area under ROC curves in Fig.~\ref{fig:results_ROC} for each algorithm configuration. Finally, we obtain the $F~scores$ by calculating a weighted combination of $precision$ and $recall$ as
\begin{equation}
F_{\beta} = (1+{\beta}^2) \frac{precision \cdot recall}{{\beta}^2 precision+recall},
\end{equation}
where $\beta$ is set to $0.5$, $1.0$, and $2.0$ to set the relative importance of $precision$ and $recall$. We report the RAPD detection performance of each algorithm configuration around their optimal operation points in Table~\ref{tab:results} in which we highlight the best performance for each evaluation metric with bold font.

\begin{figure}[htbp!]
\centering
\vspace{4mm}
\begin{minipage}[b]{\linewidth}
  \centering
\includegraphics[width=\textwidth]{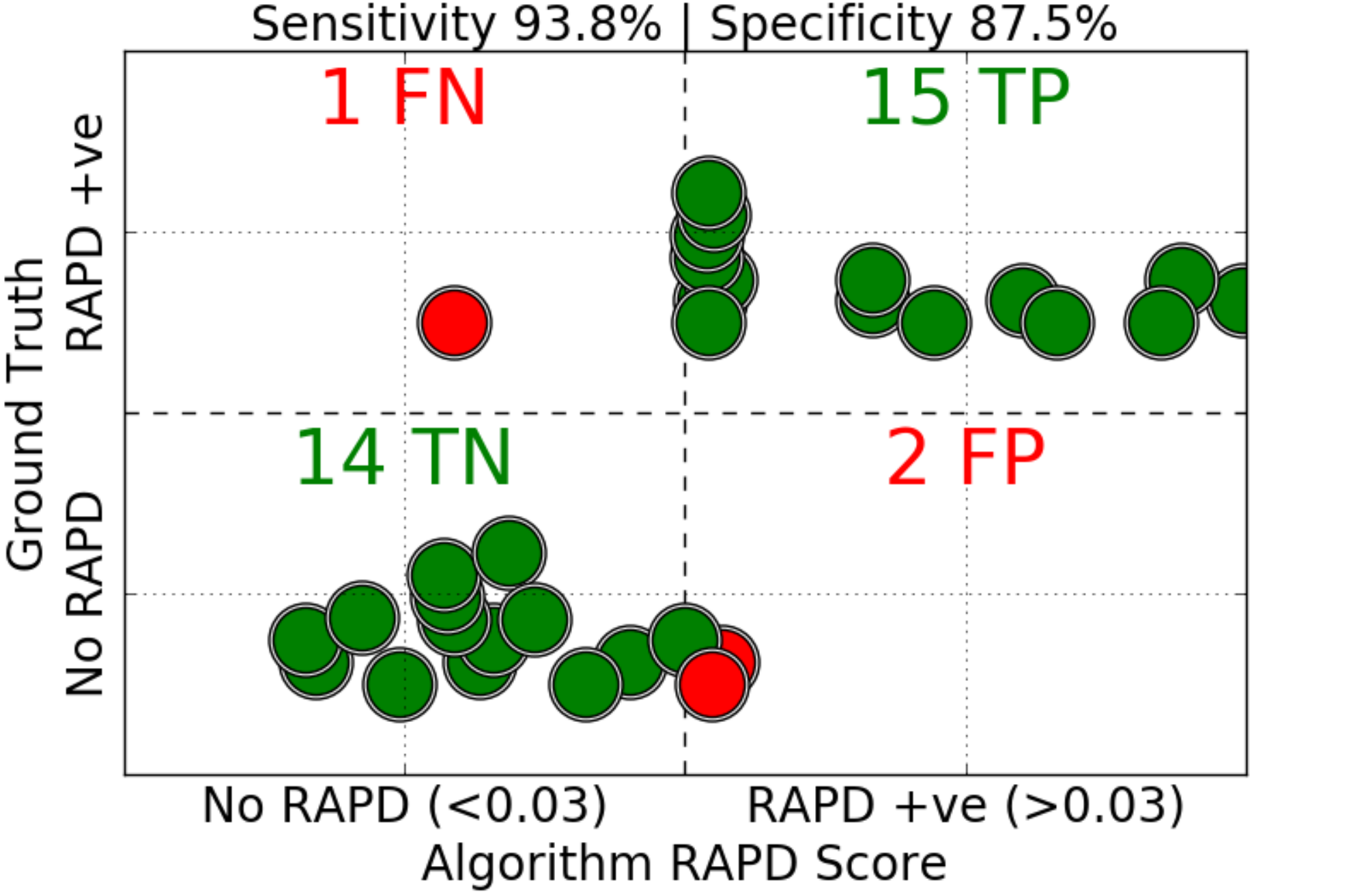}
\end{minipage}
\centering
\caption{Ground truth versus estimated RAPD conditions.}
\label{fig:results_scatter}
\end{figure}

Baseline algorithm combined with cropping and Pearson-based dissimilarity leads to highest RAPD detection performance in all evaluation categories as highlighted with a yellow background in Table~\ref{tab:results}. Based on the highest performing algorithm, we provide a scatter plot of ground truth RAPD conditions versus estimated RAPD conditions in Fig.~\ref{fig:results_scatter}. In the scatter plot, x axis corresponds to RAPD index obtained from best performing algorithm and y axis corresponds to ground truth RAPD classes. In the scatter plot, top right region corresponds to true positives and bottom left corresponds to true negatives. Top left region corresponds to false negatives and bottom right corresponds to false positives. Overall, there are only 2 false RAPD detections (FP) and 1 missed RAPD detection ($FN$) out of 32 test cases.

\section{Conclusion}
\label{sec:conc}
We automated manual swinging flashlight test with a portable imaging device and introduced an algorithm to objectively detect relative afferent pupilary defect (RAPD). Preliminary results show that introduced algorithm can correctly identify the RAPD condition of 29 cases out of 32. In this study, we only focused on the RAPD condition. However, in the future work, we plan to utilize \texttt{lab-on-a-headset} and developed algorithms for other clinical conditions related to pupillary assessment. Our objective is to develop portable systems backed with artificial intelligence that can operate in remote locations, and serve underprivileged communities and patients with mobility constraints.


\end{document}